\documentclass[a4paper,conference]{IEEEtran}
\usepackage{booktabs,makecell}
\usepackage{multirow}
\usepackage{balance}
\usepackage{times}
\usepackage{epsfig}
\usepackage{xurl}
\usepackage{amsmath}
\usepackage{amssymb}

\begin{document}
%
\title{MEG: Multi-Evidence GNN for \\Multimodal Semantic Forensics}

\author{\IEEEauthorblockN{Ekraam Sabir, Ayush Jaiswal, Wael AbdAlmageed and Prem Natarajan}
\IEEEauthorblockA{USC Information Sciences Institute\\
Marina Del Rey\\
California, USA \\
Email: \{esabir, ajaiswal, wamageed, pnataraj\}@isi.edu}
}


%


\maketitle

\begin{abstract}
Fake news often involves semantic manipulations across modalities such as image, text, location etc and requires the development of multimodal semantic forensics for its detection. Recent research has centered the problem around images, calling it image repurposing -- where a digitally unmanipulated image is semantically misrepresented by means of its accompanying multimodal metadata such as captions, location, etc. The image and metadata together comprise a multimedia package. The problem setup requires algorithms to perform multimodal semantic forensics to authenticate a query multimedia package using a reference dataset of potentially related packages as evidences. Existing methods are limited to using a single evidence (retrieved package), which ignores potential performance improvement from the use of multiple evidences. In this work, we introduce a novel graph neural network based model for multimodal semantic forensics, which effectively utilizes multiple retrieved packages as evidences and is scalable with the number of evidences. We compare the scalability and performance of our model against existing methods. Experimental results show that the proposed model outperforms existing state-of-the-art algorithms with an error reduction of up to $25\%$.
\end{abstract}


%
\IEEEpeerreviewmaketitle

\section{Introduction}\label{introduction}
Credibility of news on social media has been very low recently because of fake news \cite{schmierbach_little_2012}. The severity of the problem is indicated by its ability to affect elections, manipulate public opinion and in general, spread potentially malicious misinformation \cite{spohr_fake_2017}\cite{allcott_social_2017}. Recognizing this problem, social media platforms, such as Twitter, have conducted studies and invested in research for understanding this phenomena \cite{vosoughi_spread_2018}. The social or moral obligation to contain falsehood notwithstanding the economic and political ramifications of fake news increase the importance of developing methods to detect fake news.  The term \textit{fake news} colloquially refers to factually incorrect information. 
Technically speaking, \textit{fake news} and alternative terms such as \textit{hoax} and \textit{rumor} usually refer to manipulated multimedia (e.g. text, images, video, etc.) used for spreading incorrect information. That includes image-centric manipulations where images are digitally edited and/or associated with altered metadata.

\begin{figure}[t]
\begin{center}
   \includegraphics[width=0.8\columnwidth]{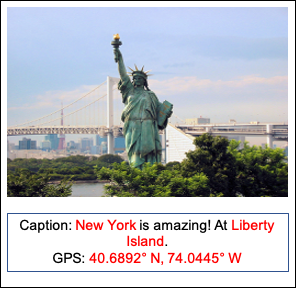}
\end{center}
   \caption{An example of multimodal semantic manipulation or image repurposing, in a multimedia package. The image was originally taken in Japan and has been used to falsify location semantics.}
\label{fig:package}
\end{figure}

Image repurposing is a relatively new research problem that deals with multimodal semantic forensics. Jaiswal et al. \cite{jaiswal_multimedia_2017} define image repurposing as the problem of semantically misrepresenting an image with falsified multimodal metadata such as captions, location, etc. The image and accompanying metadata together comprise a \textit{multimedia package}. Figure \ref{fig:package} shows an example of image repurposing presented as a multimedia package. The image was originally taken at a replica of the Statue of Liberty located at Odaiba Island in Japan, but has been repurposed to misrepresent its location. In the example, the supporting multimodal information is text and global positioning system (GPS) coordinates. The problem setup involves detecting whether a multimedia package has been semantically repurposed with the help of an external reference dataset, which is a knowledge base of packages that are assumed to contain unmanipulated information. In this setup, each package in the reference dataset is a potential evidence for verifying a query package. Sabir \textit{et al.} \cite{sabir_deep_2018-1} introduced the multimodal entity image repurposing (MEIR) dataset with challenging manipulations and presented a deep multimodal model (DMM) which achieved state-of-the-art performance on MEIR. However, a shortcoming of DMM is that it is capable of utilizing only one evidence from the reference dataset for repurposing detection. DMM does not scale to handle multiple retrieved packages and hence, does not leverage additional information for performance improvement.

In this paper, we propose MEG -- a multi-evidence graph neural network (GNN) model for multimodal semantic forensics. The GNN in our model inherently provides invariance to the order of packages retrieved from the reference dataset. This paper has the following contributions:

\begin{itemize}
    \item A new GNN-based model for multimodal semantic forensics that achieves state-of-the-art performance
    \item A scalable model for assimilating arbitrary number of evidences for semantic repurposing detection
\end{itemize}

The remainder of this paper is organized as follows: Section \ref{related_works} discusses related work. Section \ref{method} describes the proposed model. Results of our evaluation are discussed in Section \ref{evaluation}. Finally, Section \ref{conclusion} concludes the paper and provides directions for future work.

\section{Related Work}\label{related_works}

\begin{figure*}[t!]
\begin{center}
\includegraphics[width=\textwidth,keepaspectratio]{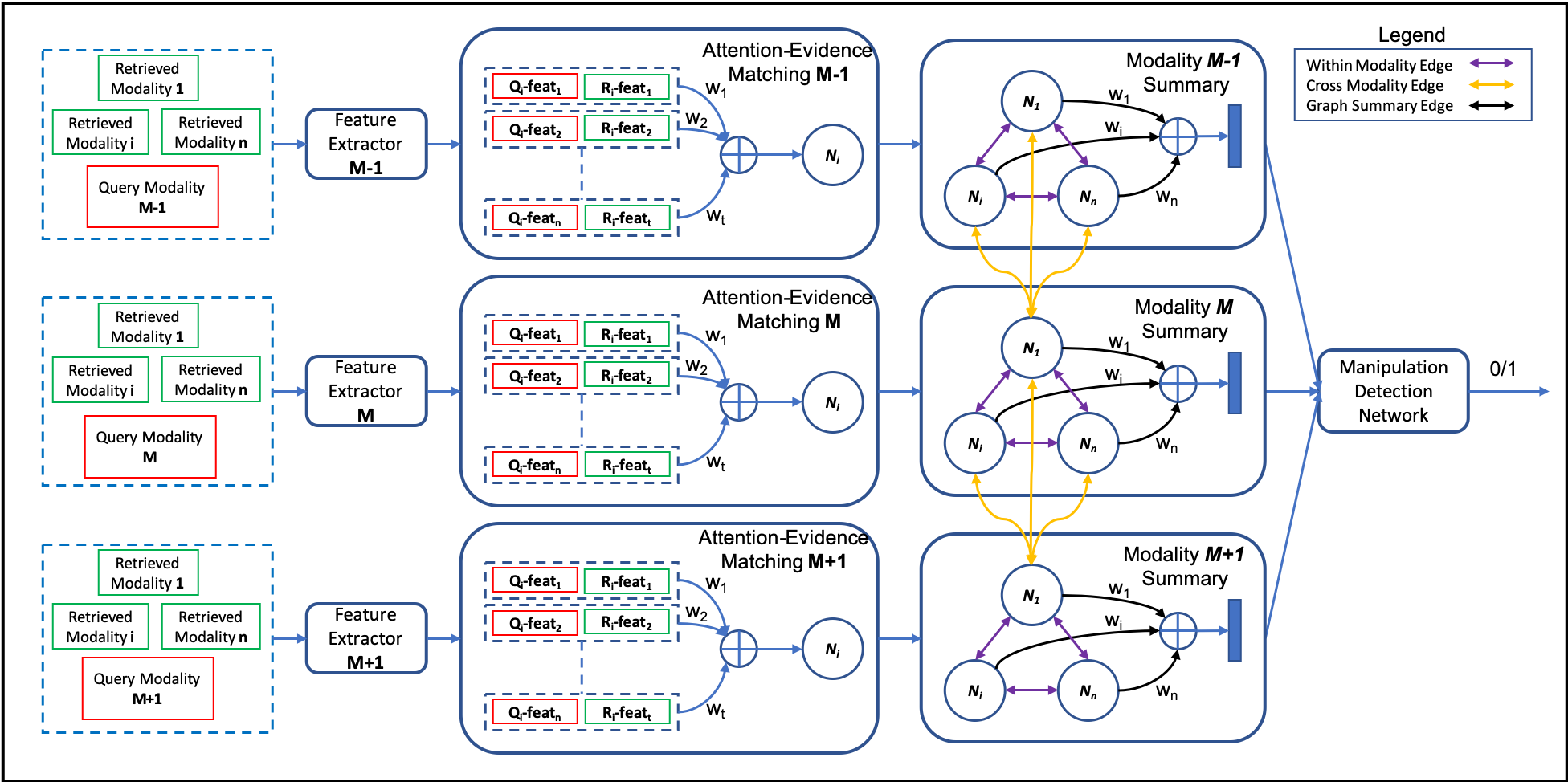}
\end{center}
   \caption{Our Model diagram. Modalities from each retrieved package are organized together and are processed through a dedicated branch of the modality. The Evidence Matching layer matches query and retrieved features side-by-side and weights it to produce a node initialization. A graph neural network is used to summarize each modality for final manipulation detection. The crossmodal connections form a complete graph in implementation, but few connections are shown for simplicity.}
\label{fig:model_diagram}
\end{figure*}

While image repurposing itself is a type of fake news, most fake news forensics do not involve images or multimodal semantics. The problem is instead tackled solely from a natural language processing (NLP) perspective. There are two popular ways to tackle the problem: (1) classification of flat feature vectors \cite{gupta_tweetcred:_2014}\cite{liu_real-time_2015} and (2) as an epidemic on a social graph \cite{wu_false_2015}\cite{jin_epidemiological_2013}. Feature vector based approaches capture information about text content such as presence of swear words, urls or information about users such as number of followers. In  \cite{gupta_tweetcred:_2014} and \cite{liu_real-time_2015} tweets are classified according to tweet content and metadata information for real time analysis. Jin \textit{et al.}\cite{jin_epidemiological_2013} investigate the use of a diffusion model to classify rumors. Wu \textit{et al.}\cite{wu_false_2015} model message propagation as a tree and use graph kernels to measure similarities for classifying tweets.

Digital image manipulation is a component of fake news, considering it can be used to misrepresent information. The connection between fake news and forged images is supported in \cite{zampoglou_web_2016}, where Zampoglou \textit{et al.} identify the need for checking images for journalism. They develop a web-based graphical user interface (GUI) for verifying images using existing algorithms. Digital image manipulation detection has been studied extensively \cite{farid_image_2009}\cite{wu_deep_2017}\cite{qureshi_bibliography_2015}. Pixel level image manipulations fall into copy-move, image-splicing, resampling and retouching categories \cite{qureshi_bibliography_2015}.

Image repurposing represents a component of fake news where images are not digitally manipulated, rather semantically misrepresented. It brings multimodal semantics into focus which has not been accounted for in previous approaches. Jaiswal \textit{et al.}\cite{jaiswal_multimedia_2017} introduced the problem of image repurposing. However the manipulations introduced in their dataset were unlikely to fool people and their reference dataset did not have directly linked evidences to query packages. They proposed a joint embedding model for images and text followed by outlier detection. This approach was unsupervised and worked on manipulations requiring common world knowledge. Sabir \textit{et al.}\cite{sabir_deep_2018-1} introduced the multimodal entity image repurposing (MEIR) dataset with more realistic manipulations. Their deep multitask (DMM) model retrieves one package from the reference dataset as evidence and uses it for repurposing/semantic manipulation detection. However, their model does not scale to multiple evidences because their multitask framework covers only one retrieved package. Jaiswal \textit{et al.} \cite{jaiswal2019aird} introduced an adversarial model for image repurposing detection, where an active counterfieter helps train the detection network. Recently Budack \textit{et al.} \cite{muller2020multimodal} attempted to detect manipulations with a web-based application that verifies cross-modal information.

Processing of sequential inputs have been well studied with recurrent neural networks \cite{sutskever_sequence_2014}. However, sequential networks may not be optimal when processing unordered inputs as shown in \cite{vinyals_order_2015}. Vinyals \textit{et al.} \cite{vinyals_order_2015} introduced a read-process-write (RPW) network for order invariant processing of set inputs. Their working mechanism is to learn a content-based representation that is invariant to input order. The read network involves learning representations for each input sample, the process network encodes all read network outputs into an order agnostic representation and the write network decodes the process network output. Application of order invariant methods is catching up, such as for object detection and multi-class image classification in \cite{rezatofighi_deepsetnet:_2017}.

Graph Neural Networks (GNNs) were introduced by Scarselli \textit{et al.}\cite{scarselli_graph_2009}. A GNN receives graph structured data as input and processes it with neural networks. Li \textit{et al.}\cite{li_gated_2016} introduced a variant of GNNs called gated graph neural networks (GG-NNs) with gated recurrent units (GRU) \cite{cho_learning_2014} for updating information between nodes. Li \textit{et al.}\cite{li_gated_2016} demonstrated the use of GG-NNs on graph reachability problems and bAbI tasks \cite{weston_towards_2015}. They have also been used to learn from knowledge graphs for image classification \cite{marino_more_2017} and for learning properties of chemical molecules \cite{gilmer_neural_2017}. There are variations of GNNs available in literature for different applications. In \cite{li_gated_2016}, the authors propose another architecture - gated graph sequence neural networks (GGS-NNs), which is a sequence of GG-NNs producing intermediate outputs. Graph convolutional networks (GCNs) \cite{duvenaud_convolutional_2015}\cite{niepert_learning_2016} are inspired from convolutional neural networks (CNNs) and learn local receptive fields on graph structured input. In the context of multimodal semantic forensics, GNN provides a way to assimilate multiple evidences.

\section{Multimodal Semantic Forensics}\label{method}

As discussed in Section \ref{introduction}, reference datasets in image repurposing problems often have multiple evidences for verification of the query package. However, previous methods for image repurposing detection cannot handle variable number of retrieved packages \cite{jaiswal_multimedia_2017}\cite{sabir_deep_2018-1}. This shortcoming prevents these models from leveraging potentially multiple instances of related packages for performance improvement. As such, a driving motivation of our model design is to make it scalable to multiple retrieved packages. Additionally, as discussed in \cite{sabir_deep_2018-1}, it is possible for a modality to be missing from a package or the dataset itself. For example an image may be accompanied by text or location information or both. Under such circumstances, it is also important to keep the model flexible to handle an arbitrary number of modalities. In order to ensure this, our model processes each modality in a different branch which is architecturally the same. Effectively, each branch has the same architecture, but with different learned weights for each modality. Each branch has three major components - (1) feature extraction, (2) evidence matching and (3) modality summary. They are preceded by package retrieval for evidences and followed by the manipulation detection layer. Figure \ref{fig:model_diagram} gives an overview of our model. We describe the motivation and design of each component of our model below.

\subsection{Package Retrieval}\label{package_retrieval}

Verification of a query package requires additional information from a reliable reference dataset. Since packages retrieved from the reference dataset form the basis for authentication, the package retrieval system is an important component of the overall method. We use a package retrieval system similar to \cite{sabir_deep_2018-1}. In \cite{sabir_deep_2018-1}, the authors score each modality of the query package against the corresponding modality of all packages in the reference dataset. A reference package with the top-1 combined score across all modalities is retrieved. We extend this method to retrieve $k$ packages with the highest scores. 

\subsection{Feature Extraction}\label{feature_extraction}

Learned feature extraction is an important component of deep learning models. Previous literature in image repurposing detection has used pretrained models to extract features from all modalities of a package. We follow a similar approach using convolutional neural network (CNN) based models, word2vec and global positioning system (GPS) coordinates for image, text and location feature extraction respectively. Specific details on models used for feature extraction are discussed in Section \ref{datasets}.

\subsection{Attention-based Evidence Matching}\label{evidence_matching}

Semantic forensics can involve a subtle but specific change of detail which can be hard to detect at a glance. Additionally, the information (entity, location, etc.) involving both the manipulation and evidence in query and retrieved packages is unlikely to be previously seen. It is therefore prudent to develop a method that can compare a previously unseen instance of manipulation and evidence without memorizing it. This requirement is in contrast to classical computer vision models that reward memorization of training examples such as associating the word \textit{dog} with a corresponding image in a standard classification task. We address the problem of dealing with previously unseen manipulations and evidences with an attention-based evidence matching module, shown in Figure \ref{fig:model_diagram}. Evidence matching compares concatenated query and retrieved features with a soft attention mechanism for selecting important matches. For query and retrieved features $q$ and $r$ respectively, a concatenated feature vector $[q,r]$ is processed with 1D convolution network (CNN) for matching. A soft attention model on top of the concatenated representation, followed by a dense layer to compute a matched feature $feat$. This layer can be represented by Equation \ref{eq:attention-feat}
\begin{equation}\label{eq:attention-feat}
    feat = FC\bigg(\sigma\big(Conv([q,r])\big)\odot([q,r])\bigg)
\end{equation}
where $Conv$ is a 1D CNN, $\sigma$ is soft attention and $FC$ is a dense layer for dimensionality matching across modalities.

\subsection{Modality Summary}\label{modality_summary}

The retrieved packages represent a \emph{bag-of-packages} without specific order. We use a graph neural network (GNN) for each modality that considers all possible comparisons between the query and retrieved packages.  The graph network makes the overall system (1) flexible enough to scale to an arbitrary number of retrievals and (2) invariant to the order of retrieved packages. Each node in the graph network is updated with respect to its adjacent nodes allowing simultaneous updates. The graph is then  summarized into one modality-level representation. Each node contributes directly to the final graph summary. This is different from recurrent networks where the latent embedding is updated in sequence, making them order-dependent. A node $v$ in a GNN is represented by the hidden state $h_v$. A forward pass through a GNN is divided into propagation and output steps. The propagation step updates nodes along edges in the graph for $T$ timesteps. It can be thought of as a gated recurrence along paths in the graph, similar to long short-term memory network (LSTM) recurrence. The output step produces a graph level vector representation by combining hidden states of nodes with an attention mechanism. The model is summarized by Equations \ref{eq:graph1}-\ref{eq:graph6}
\begin{equation}\label{eq:graph1}
h_v^1 = [x_v, 0]^T
\end{equation}
\begin{equation}\label{eq:graph2}
a_v^{(t)} = A_v^T[h_1^{(t-1)} ...h_N^{(t-1)}]^T + b
\end{equation}
\begin{equation}\label{eq:graph3}
z_v^t = \sigma(W^za_v^{(t)} + U^zh_{v}^{(t-1)})
\end{equation}
\begin{equation}\label{eq:graph4}
r_v^t = \sigma(W^ra_v^{(t)} + U^rh_v^{(t-1)})
\end{equation}
\begin{equation}\label{eq:graph5}
\widetilde{h_v^{(t)}} = tanh(Wa_v^{(t)} + U(r_{v}^{t} \odot h_v^{(t-1)}))
\end{equation}
\begin{equation}\label{eq:graph6}
h_v^{(t)} = (1-z_v^t) \odot h_v^{(t-1)} + z_v^t \odot \widetilde{h_v^{(t)}}
\end{equation}

The hidden state is initialized with an initial representation $x_v$ according to the application and padded with 0 to match dimensions if needed. $A$ is the adjacency matrix and $a_v$ is the summation of adjacent node embeddings based on edge type. Equations \ref{eq:graph3}-\ref{eq:graph6} represent updates using a GRU. The graph neural network effectively summarizes the potential for manipulation in a learned representation for each modality. We use a complete graph (adjacency matrix of ones except along the diagonal) with one timestep of propagation. This allows simultaneous update of all nodes throughout the graph in an order agnostic manner. The final graph output $G_m$ for modality $m$ of our model is a weighted average of activation all $N$ nodes as shown in Equation \ref{eq:graph_output}:

\begin{equation}\label{eq:graph_output}
    G_{m} = \sum_{v=1}^{N}\big(h_v^1 \odot Att(h_v^1)\big)
\end{equation}


The weights are estimated using a neural network $Att$. Since the model is set up for variable number of inputs, the scale of adjacent node embeddings $a_v$ may fluctuate by an order of magnitude. To control for the variation, we modify Equation \ref{eq:graph2} by scaling it down by the number of adjacent nodes. For our fully connected graph setup, modality $m$ with $N$ nodes has $N-1$ adjacent nodes, as shown in  \ref{eq:graph2}.
\begin{equation}\label{eq:graph7}
    a_v^{(t)} = \frac{A_v^T[h_1^{(t-1)} ...h_N^{(t-1)}]^T + b}{N-1+\epsilon}
\end{equation}

This summarizes each modality into a single graph output, with nodes of the same modality. However, it has been shown that cross-modal learning helps with multimodal tasks \cite{ngiam_multimodal_2011}. To incorporate cross-modal learning into our model we add cross-modal graph connections. The adjacency matrix is expanded to include nodes from adjacent modalities. We validate the performance of cross-modal connections later in Section \ref{results}. For $m$ modalities, each with $N_{m}$ nodes, a general update to Equation \ref{eq:graph7} for arbitray nodes in adjacent modalities is shown in Equation \ref{eq:graph8}.
\begin{equation}\label{eq:graph8}
    a_v^{(t)} = \frac{A_v^T[h_1^{(t-1)} ...h_N^{(t-1)}]^T + b}{N_i-1+\sum\limits_{j=1,j\neq i}^{m} N_{j}+\epsilon}
\end{equation}
However, considering that each modality generates an equal number of nodes and fully-connected cross-modal edges are used, Equation \ref{eq:graph8} is simplified to Equation \ref {eq:graph9}.
\begin{equation}\label{eq:graph9}
    a_v^{(t)} = \frac{A_v^T[h_1^{(t-1)} ...h_N^{(t-1)}]^T + b}{m*N-1+\epsilon}
\end{equation}
The $\epsilon$ term takes care of zero adjacency for each node in a graph. Finally, the output representation for each modality is combined as described next.

\subsection{Manipulation Detection}\label{manipulation_detection}

A feed-forward network on top of concatenated modality summary outputs is used for the final manipulation detection. This layer combines all branches of modalities into a single binary prediction. 

\subsection{Implementation Details}\label{implementation_details}

Our model was implemented in Keras and trained with ADAM optimizer with an initial learning rate of $0.001$. All parameters had default values, unless otherwise mentioned. All edge layers and feedforward layers have ReLU activation function. We trained all our models with a batch size of $32$ and subsampled models within each epoch for selecting the best model.

\section{Evaluation}\label{evaluation}
This section describes benchmarks in Section \ref{datasets} and both quantitative and qualitative results in Section \ref{results}.

\subsection{Benchmark Datasets}\label{datasets}
We perform experimental evaluation on MEIR \cite{sabir_deep_2018-1}, which is the most challenging dataset for image repurposing detection. We also evaluate on \textit{Google Landmarks} \cite{noh_large-scale_2017} and \textit{Painter by Numbers} \cite{noauthor_painter_nodate} datasets which were originally released for different tasks, but can be adapted for semantic forensics. The adapted splits for Google Landmarks and Painter by Numbers used in \cite{jaiswal2019aird} have repeated locations and painters in training and testing. Keeping in line with the idea in \cite{sabir_deep_2018-1} that manipulations are unseen in training and test, we adapt a new split with mutually exclusive manipulations in training and test set. 

\paragraph{MEIR:} It is a multimodal dataset comprising images, text and location modalities. Manipulations are present in text and location modalities and comprise three types of entity manipulations --- person, location and organization. Manipulations are also coherent within a package i.e. a location manipulation within text will result in corroborating manipulations in GPS coordinates. The dataset comprises 82,156 packages in reference dataset and 57,940 packages split between training, test and validation sets. It should be noted that all packages in training, test and validation conform to different events. This helps in evaluating the generalizability of models to unseen semantic manipulations. 

\paragraph{Google Landmarks:} We use Google Landmarks for further evaluating location manipulation, since locations can easily confuse people, especially if the landmark in the photo is not well recognized by the person. This is one of the manipulations present in MEIR, but is mixed with all other manipulations. This dataset is available as a part of a Kaggle competition\footnote{https://www.kaggle.com/google/google-landmarks-dataset/home}. The modified task for semantic forensics on this dataset is to identify if the landmark associated with a query package is correct. The complete dataset is extremely large with over 1.2 million images and 14,951 different landmarks. We prune the dataset, keeping landmarks with at least 3 images and at most 50 images. This leaves us with 152,074 images split into 78,573 images for reference and 73,501 images for train, test and validation which is further split in a 70-10-20 ratio. We create believable semantic manipulations by swapping similar images. Images are determined to be similar using a kd-tree search. During test, we ensure that landmarks in training, test and validation form a disjoint set. This ensures that the model is robust in identifying unseen landmarks. Image features are generated using NetVLAD \cite{arandjelovic_netvlad:_2016}, followed by principal component analysis (PCA) and $l_2$ normalization as used in \cite{jaiswal2019aird}. Landmark IDs are encoded using 50 dimensional random embeddings. We also measure our retrieval accuracy using mean average precision (MAP). With this scheme of manipulation and feature embedding, a cosine similarity based retriever as described in \cite{sabir_deep_2018-1}, but generalized to 5 package retrieval achieves 0.81 mean average precision.

\paragraph{Painter by Numbers:} We evaluate the proposed system on painting forgeries, which is an old problem with counterfeits being created for paintings by famous artists. This is a high stakes problem with art experts being called in to validate paintings. We create a painting repurposing dataset from the \textit{Painters by Numbers} dataset. This dataset was also released as a part of a Kaggle challenge\footnote{https://www.kaggle.com/c/painter-by-numbers}. We restructure the dataset for semantic forensics, where the identity of the artist for a given painting is potentitally manipulated. After ensuring that each artist has at least three images, the dataset is split into 36,669 reference images, and 36,164 images for train, test and validation. There are 1000 different artists in the dataset. To create manipulations, we use a kd-tree for finding similar paintings and swap the artists. There is no overlap between artists in training, test and validation, to ensure generalization. Image features are extracted using the winner's model from the competition\footnote{https://github.com/inejc/painters}, similar to \cite{jaiswal2019aird}. For painter IDs, we generate 50 dimensional random embeddings. Again, using the retrieval scheme in \cite{sabir_deep_2018-1}, for top-5 packages, we achieve 0.72 mean average precision.

\subsection{Evaluation Results}\label{results}

We use accuracy, area under the receiver operating characteristic curve (AUC), and F\textsubscript{1}-clean and F\textsubscript{1}-tampered (F\textsubscript{1} scores for unmanipulated and manipulated class respectively) scores as evaluation metrics. Previous works have used these metrics \cite{sabir_deep_2018-1}\cite{jaiswal_multimedia_2017}. We perform ablation experiments to test the scalability and order invariance of our model. A summary of the results is discussed. We also evaluate our model on benchmark datasets and discuss the quantitative and qualitative results.

\begin{table}[t!]
    \begin{center}
    \setlength{\tabcolsep}{0.2em} 
    \caption{Ablation experiments for verifying scalability. We replace the modality summary (GNN) component of MEG with other models. All variants are trained on two and tested on five packages. AUC scores are reported.}
    \label{table_robust}
    \begin{tabular}{c|c|c|c}
        \hline
         \textbf{Ablation} &  \textbf{Train on} & \textbf{Test on} & \textbf{Relative} \\
         \textbf{Model} & \textbf{2 Packages} & \textbf{5 Packages} & \textbf{Drop}\\
         \hline
         MEG (Ours) & \textbf{0.91} & \textbf{0.91} & \textbf{0\%} \\
         \hline
         MEG - GNN + GRU & 0.90 & 0.88 & 20\% \\
         MEG - GNN + LSTM & 0.90 & 0.85 & 50\% \\
         MEG - GNN + RPW & 0.89 & 0.89 & 0\% \\
         MEG - scaling & 0.90 & 0.85 & 50\% \\
         \hline
    \end{tabular}
    \end{center}
\end{table}

\paragraph{Scalability:} A contribution of our model is the ability to handle variable number of related packages. The modality summary module of our model is responsible for providing scalability. Keeping this in mind, we perform two categories of ablation experiments as shown in Table \ref{table_robust}: (1) replacing the modality summary network with standard recurrent networks (GRU and LSTM) and the read-process-write (RPW) network from \cite{vinyals_order_2015} (2) removing scaling modifications we made to the graph network in Section \ref{modality_summary}. For this set of experiments we train our model for up to two packages and test on 5 packages. A drop in performance at 5 packages indicates that the model does not scale. We train all models with a minimum of two packages to avoid an empty adjacency matrix for GNN. The results clearly support the scalability of the proposed model.

\begin{table}[t!]
    \begin{center}
    \caption{Ablation experiments for verifying order invariance. We replace the modality summary (GNN) in MEG with other models. Before and After columns show performance before and after reversing retrieval order. AUC scores reported.}
    \label{table_order}
    \begin{tabular}{c|c|c|c}
        \hline
         \textbf{Ablation} &  \multirow{2}{*}{\textbf{Before}} & \multirow{2}{*}{\textbf{After}} & \textbf{Relative} \\
         \textbf{Model} & & & \textbf{Drop} \\
         \hline
         MEG (Ours) & \textbf{0.92} & \textbf{0.92} & \textbf{0.0\%} \\
         \hline
         MEG - GNN + GRU & 0.91 & 0.83 & 88.8\% \\
         MEG - GNN + LSTM & 0.91 & 0.87 & 44.4\% \\
         MEG - GNN + RPW & 0.89 & 0.89 & 0.0\% \\
        \hline
    \end{tabular}
    \end{center}
\end{table}

\paragraph{Order Invariance:} A result of the GNN based modality summary module in our model is invariance to input ordering. We perform ablation experiment by replacing the modality summary module with standard recurrent networks (GRU and LSTM) and an existing order agnostic model - read-process-write (RPW) network from \cite{vinyals_order_2015}. It has been reported that recurrent networks suffer from order dependence issues, resulting in performance drop for input order changes between training and test \cite{vinyals_order_2015}. We train our model for 5 packages and test by reversing the training order. A drop in performance indicates that the model variation is not model invariant. Results are presented in Table \ref{table_order}. It is evident that LSTM or GRU based variations of our model are not order invariant. As expected, replacing GNN with RPW \cite{vinyals_order_2015} in the modality summary layer maintains order invariance, but leads to a performance drop.

\begin{table}[t!]
    \begin{center}
    \setlength{\tabcolsep}{0.4em} 
    \caption{Summary of our ablation experiments. MEG outperforms across all three factors. AUC scores are reported.}
    \label{tab_summary}
    \begin{tabular}{c|c|c|c}
        \hline
         \textbf{Ablation} & \textbf{Order} & \multirow{2}{*}{\textbf{Scalability}} & \multirow{2}{*}{\textbf{Score}} \\
         \textbf{Model} & \textbf{Invariance} & & \\
         \hline
         MEG (Ours) & \checkmark & \checkmark & \textbf{0.92}\\
         \hline
         MEG - GNN + GRU & & & 0.91 \\
         MEG - GNN + LSTM & & & 0.91 \\
         MEG - GNN + RPW & \checkmark & \checkmark & 0.89 \\
         MEG - scaling & \checkmark & & 0.90 \\
         \hline
    \end{tabular}
    \end{center}
\end{table}

\paragraph{Ablation Summary:} We summarize the ablation results in Table \ref{tab_summary}. Three properties are considered: scalability, order invariance and detection performance. Our method satisfies all properties while maintaining best performance for all comparisons.

\begin{table*}[t]
\begin{center}
\setlength{\tabcolsep}{1.2em} 
\caption{Performance of our proposed model (MEG) against existing methods from \cite{sabir_deep_2018-1} across all three benchmark datasets.}
\label{table_performance}
\begin{tabular}{c|ccc|ccc|ccc}
\hline
\multirow{2}{*}{\textbf{Metric}} & \multicolumn{3}{c|}{\textbf{MEIR}} & \multicolumn{3}{c|}{\textbf{Painter by Numbers}} & \multicolumn{3}{c}{\textbf{Google Landmarks}} \\
\cline{2-10}
 &  \textbf{SRS} & \textbf{DMM} & \textbf{MEG} & \textbf{SRS} & \textbf{DMM} & \textbf{MEG} & \textbf{SRS} & \textbf{DMM} & \textbf{MEG}\\
\hline
F\textsubscript{1}-clean     & 0.51 & 0.80 & \textbf{0.84} & 0.70 & 0.59 & \textbf{0.83} & 0.82 & \textbf{0.87} & \textbf{0.87} \\
F\textsubscript{1}-tampered     & 0.66 & 0.80 & \textbf{0.84} & 0.80 & 0.67 & \textbf{0.79} & 0.86 & \textbf{0.87} & \textbf{0.87} \\
Accuracy     & 0.60 & 0.80 & \textbf{0.84} & 0.76 & 0.63 & \textbf{0.82} & 0.84 & \textbf{0.88} & 0.87 \\
AUC     & 0.67 & 0.88 & \textbf{0.92} & 0.77 & 0.74 & \textbf{0.86} & 0.93 & 0.93 & \textbf{0.94} \\
\hline
\end{tabular}
\end{center}
\end{table*}

\begin{table}[t!]
    \centering
    \setlength{\tabcolsep}{1.45em} 
    \caption{We measure the average number of correctly retrieved packages out of top-5 retrievals for correctly classified (True Positive and True Negative) and misclassified (False Positive and False Negative) query packages. Package retrieval accuracy positively affects final model performance.}
    \label{table_retrieval}
    \begin{tabular}{c|c|c}
        \hline
         \multirow{2}{*}{\textbf{Dataset}} &  \multicolumn{2}{c}{\textbf{Classification Category}} \\
        \cline{2-3}
         & TP+TN & FP+FN \\
         \hline
         MEIR & \textbf{3.05} & 2.60 \\
         Painter by Numbers & \textbf{3.25} & 1.00 \\
         Google Landmarks & \textbf{3.05} & 2.15 \\
         \hline
    \end{tabular}
\end{table}

\paragraph{Performance:} We compare performance against previous methods from \cite{sabir_deep_2018-1} - namely the deep multimodal model (DMM) and the semantic retrieval system (SRS). DMM is a deep learning based model which verifies a query package using top-1 retrieved package. SRS is a non-learning method which computes the Jacardian index on packages retrieved by individual modalities. It's performance is known to scale with the correctness of retrievals. Our model improves upon state-of-the-art performance across all three datasets as shown in Table \ref{table_performance}.

\begin{figure}[t]
    \centering
    \includegraphics[width=0.8\columnwidth,keepaspectratio]{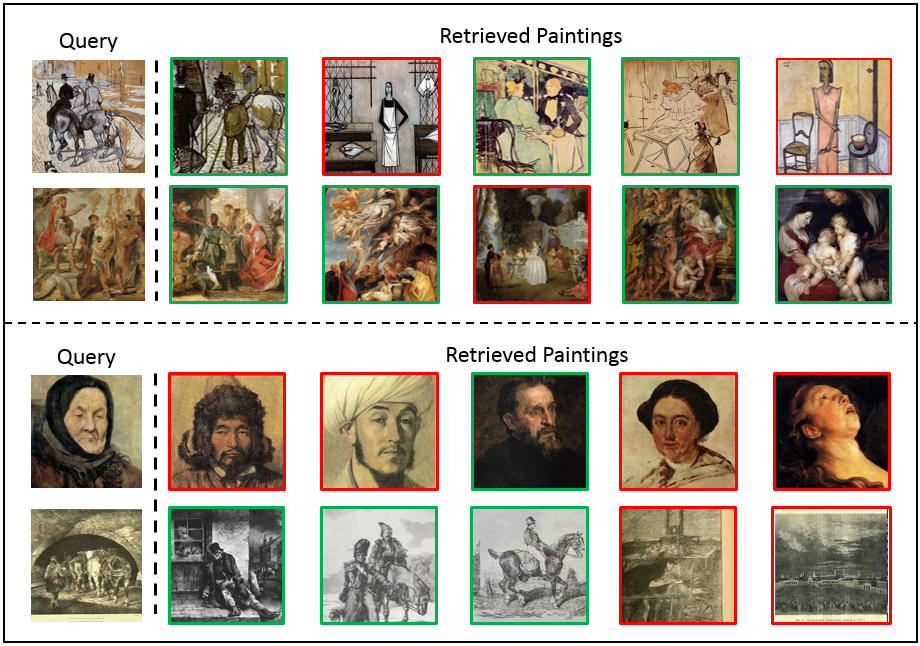}
    \caption{The top two rows contain true positive examples and the bottom two rows contain false positive samples. Across both cases it is noticeable that the repurposing/manipulation is believable. In the bottom row, the correct retrievals are visually different from the query, leading to false alarms. Green and red borders indicate correct and incorrect retrievals respectively.}
    \label{fig:painters}
\end{figure}

\begin{figure}[t]
    \centering
    \includegraphics[width=0.8\columnwidth,keepaspectratio]{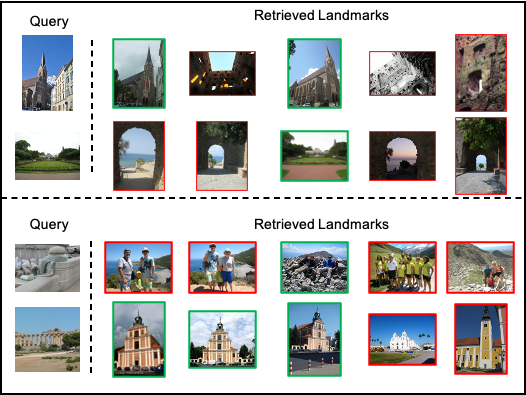}
    \caption{The top two and bottom two rows contain true positive and false positive samples respectively. In the bottom row, the correct retrievals look significantly different, leading to a false alarm. Green and red borders indicate correct and incorrect retrievals respectively.}
    \label{fig:landmarks}
\end{figure}

\begin{figure*}[t!]
    \centering
    \includegraphics[width=0.7\textwidth,keepaspectratio]{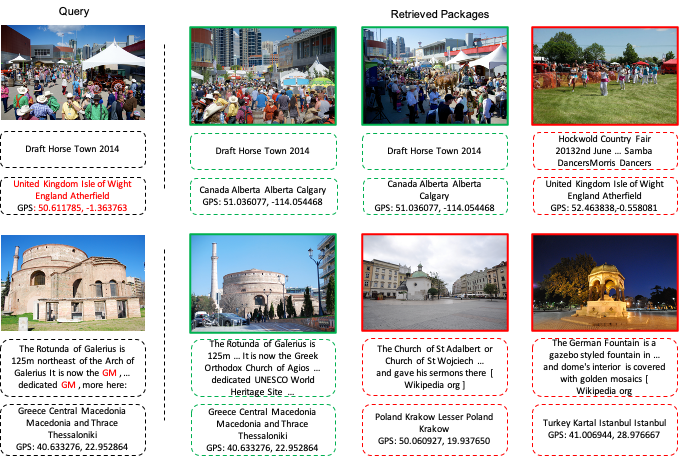}
    \caption{The two rows show true positive samples of our model. The first and second package have location and organization manipulation respectively. Green and red borders indicate correct and incorrect retrievals respectively. Metadata highlighted in red in query package is manipulation.}
    \label{fig:meir_tp}
\end{figure*}

\begin{figure*}[t!]
    \centering
    \includegraphics[width=0.7\textwidth,keepaspectratio]{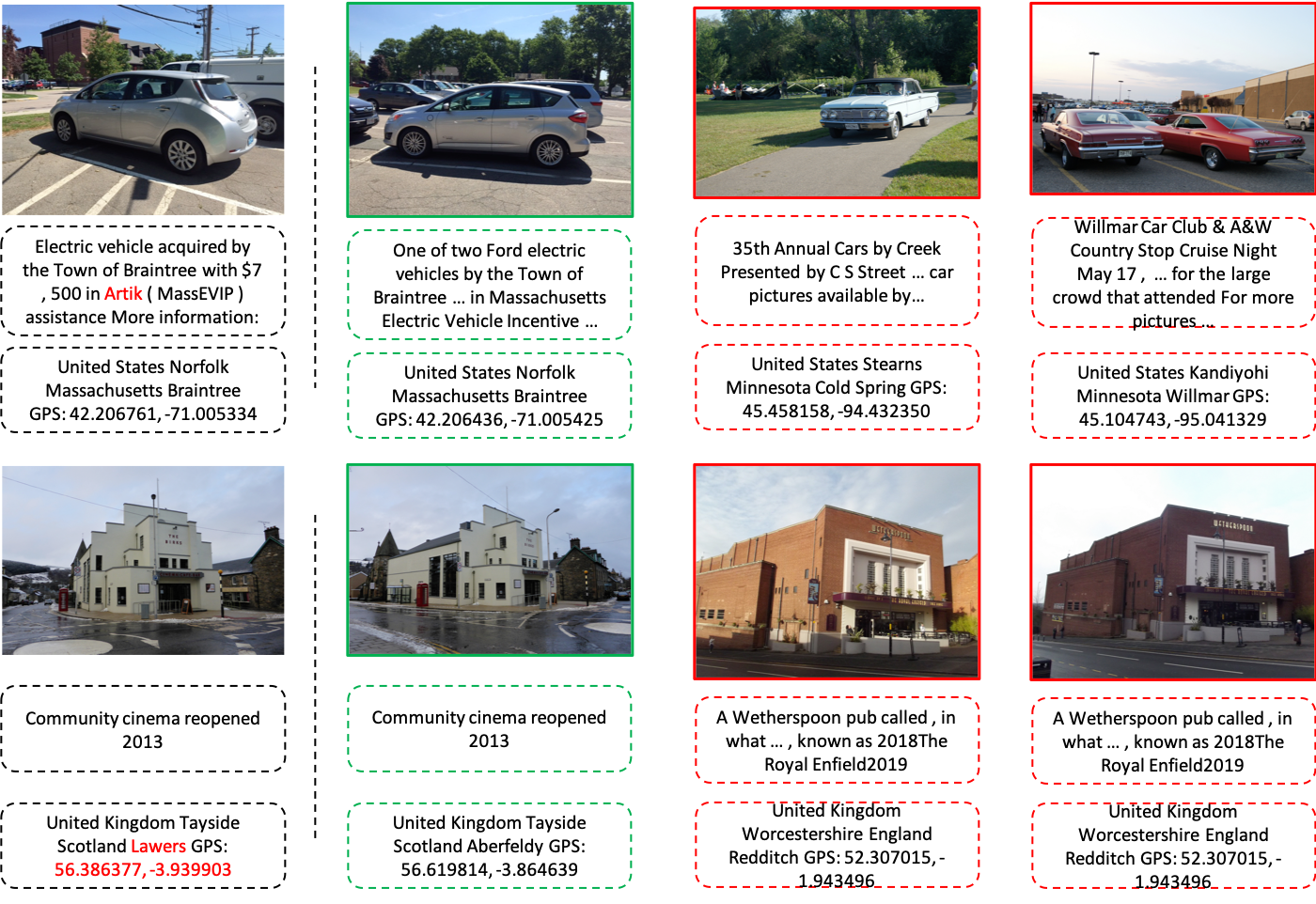}
    \caption{The two rows show false negative samples of our model. The first and second package have location and organization manipulation respectively. Green and red borders indicate correct and incorrect retrievals respectively. Metadata highlighted in red in query package is manipulation.}
    \label{fig:meir_fn}
\end{figure*}

\paragraph{Analysis:} Examples from Painters by Numbers and Google Landmarks datasets are shown in Figure \ref{fig:painters} and \ref{fig:landmarks} respectively. True positive and false negative examples from MEIR are shown in Figure \ref{fig:meir_tp} and \ref{fig:meir_fn} respectively. It is visible from the results that image repurposing performance is dependent on package retrieval performance. To further test this hypothesis, we compare the average number of correct packages retrieved between successful (true positive and true negative) and unsuccessful (false positive and false negative) classifications. The results in Table \ref{table_retrieval} show a consistently better retrieval for all correctly classified packages.

%
%

\section{Conclusion and Future Work}\label{conclusion}

Image repurposing detection is an important but emerging research area for multimodal semantic forensics and fake news detection. We presented a multi-evidence GNN model (MEG) for multimodal semantic forensics that improves upon previous state-of-the-art across three benchmark datasets. Our scaling modifications over a standard GNN make the proposed model scalable to multiple retrieved packages. Our model is order invariant compared to standard recurrent architectures. 

Besides the improvements to image repurposing detection in this paper, there are still unexplored problems remaining. MEG does not localize the exact manipulation. While successful manipulation detection can alert users to semantic manipulations, successful localization can help users reason about manipulations. Another possible area to explore is real-time multimodal semantics i.e. using the web instead of a reference dataset. These directions are left for future work.






%
\balance
{
\small
\bibliographystyle{IEEEtran}
\bibliography{Mac_bib}
}

\end{document}